\documentclass[conference]{IEEEtran}
\IEEEoverridecommandlockouts
\usepackage{cite}
\usepackage{amsmath,amsfonts,amsthm}
\usepackage{mathtools}
\usepackage{textcomp}

\usepackage{latexsym}
\usepackage{xspace}
\usepackage{xcolor,graphicx}
\usepackage{verbatim}
\usepackage{changepage} 
\usepackage{transparent}
\usepackage{algorithm}
\usepackage[noend]{algpseudocode}
\usepackage{todonotes}
\usepackage{framed,multirow}
\usepackage[normalem]{ulem}
\usepackage{url}
\usepackage{hyperref}
\usepackage{pgfplots}
\usepackage{pgfplotstable}
\usepackage{enumitem}
\usepackage{amssymb,mathtools}
\usepackage{textcomp}

\pgfplotsset{compat=1.7}
\usepackage{balance}
\usepackage[utf8]{inputenc}
\usepackage[T1]{fontenc}
\usepackage{soul}

\newcommand{\conffive}[3]{
  \begin{tikzpicture}
 \foreach \x in {0,...,4}
  {
  \draw[very thick]  (360/5*\x +90:6mm) circle (2mm) ;
  }

  \foreach \x in #1
    \fill[black]  (360/5*\x +90:6mm) circle (1mm);

 \ifnum #3>-1
   \fill[black] (360/5*#3 +90:6mm) circle (2mm);
 \fi

\foreach \x/\y in #2
  {
  \ifnum \x<0
    \draw[->,very thick] (360/5*-\x +90 +10:10mm+\y) -- (360/5*-\x 
+90-10:10mm+\y)
  \else
     \draw[->,very thick] (360/5*\x +90-10:10mm+\y) -- (360/5*\x +90+10:10mm+\y)
  \fi;
  }
  \end{tikzpicture}
} 

\newcommand{\confx}[3]{
  \begin{tikzpicture}
 \foreach \x in {0,...,6}
  {
  \draw[very thick]  (360/7*\x +90:6mm) circle (2mm) ;
  }

  \foreach \x in #1
    \fill[black]  (360/7*\x +90:6mm) circle (1mm);

 \ifnum #3>-1
   \fill[black] (360/7*#3 +90:6mm) circle (2mm);
 \fi

\foreach \x/\y in #2
  {
  \ifnum \x<0
    \draw[->,very thick] (360/7*-\x +90 +10:10mm+\y) -- (360/7*-\x 
+90-10:10mm+\y)
  \else
     \draw[->,very thick] (360/7*\x +90-10:10mm+\y) -- (360/7*\x +90+10:10mm+\y)
  \fi;
  }
  \end{tikzpicture}
} 

\newcommand{\conf}[3]{
  \begin{tikzpicture}
 \foreach \x in {0,...,6}
  {
  \draw[very thick]  (360/7*\x +90:6mm) circle (2mm) ;
  }

  \foreach \x in #1
    \fill[black]  (360/7*\x +90:6mm) circle (1mm);

 \ifnum #3>-1
   \fill[black] (360/7*#3 +90:6mm) circle (2mm);
 \fi

\foreach \x/\y in #2
  {
  \ifnum \x<0
    \draw[->,very thick] (360/7*-\x +90 +10:10mm+\y) -- (360/7*-\x +90-10:10mm+\y)
  \else
     \draw[->,very thick] (360/7*\x +90-10:10mm+\y) -- (360/7*\x +90+10:10mm+\y)
  \fi;
  }
  \end{tikzpicture}
} 

\newcommand{\namedconf}[3]
{
\begin{tikzpicture}
 \foreach \x in {0,...,6}
  {
  \draw[very thick]  (360/7*\x +90:6mm) circle (2mm) ;
  }

  \foreach \x in #1
    \fill[black]  (360/7*\x +90:6mm) circle (1mm);

 \ifnum #3>-1
   \fill[black] (360/7*#3 +90:6mm) circle (2mm);
 \fi

\foreach \x/\y in #2
    \draw (360/7*\x +90:10mm) node {$\y$};
\end{tikzpicture}
}

\newcommand{\namedconffivemult}[3]
{
\begin{tikzpicture}
 \foreach \x in {0,...,4}
  {
  \draw[very thick]  (360/5*\x +90:6mm) circle (2mm) ;
  }

  \foreach \x in #1
    \fill[black]  (360/5*\x +90:6mm) circle (1mm);

   \foreach \x in #3
     \fill[black] (360/5*\x +90:6mm) circle (2mm);

\foreach \x/\y in #2
    \draw (360/5*\x +90:10mm) node {$\y$};
\end{tikzpicture}
}

\newcommand{\namedconffive}[3]
{
\begin{tikzpicture}
 \foreach \x in {0,...,4}
  {
  \draw[very thick]  (360/5*\x +90:6mm) circle (2mm) ;
  }

  \foreach \x in #1
    \fill[black]  (360/5*\x +90:6mm) circle (1mm);

   \ifnum #3>-1
   \fill[black] (360/5*#3 +90:6mm) circle (2mm);
   \fi

\foreach \x/\y in #2
    \draw (360/5*\x +90:10mm) node {$\y$};
\end{tikzpicture}
}

\newcommand{\namedconffour}[3]
{
\begin{tikzpicture}
 \foreach \x in {0,...,3}
  {
  \draw[very thick]  (360/4*\x +90:6mm) circle (2mm) ;
  }

  \foreach \x in #1
    \fill[black]  (360/4*\x +90:6mm) circle (1mm);

   \ifnum #3>-1
   \fill[black] (360/4*#3 +90:6mm) circle (2mm);
   \fi

\foreach \x/\y in #2
    \draw (360/4*\x +90:10mm) node {$\y$};
\end{tikzpicture}
}

\newcommand{\namedconfnine}[3]
{
\begin{tikzpicture}
 \foreach \x in {0,...,8}
  {
  \draw[very thick]  (360/9*\x +90:6mm) circle (2mm) ;
  }

  \foreach \x in #1
    \fill[black]  (360/9*\x +90:6mm) circle (1mm);

 \ifnum #3>-1
   \fill[black] (360/9*#3 +90:6mm) circle (2mm);
 \fi

\foreach \x/\y in #2
    \draw (360/9*\x +90:10mm) node {$\y$};
\end{tikzpicture}
}

\newcommand{\confnine}[3]{
  \begin{tikzpicture}
 \foreach \x in {0,...,8}
  {
  \draw[very thick]  (360/9*\x +90:6mm) circle (2mm) ;
  }

  \foreach \x in #1
    \fill[black]  (360/9*\x +90:6mm) circle (1mm);

 \ifnum #3>-1
   \fill[black] (360/9*#3 +90:6mm) circle (2mm);
 \fi

\foreach \x/\y in #2
  {
  \ifnum \x<0
    \draw[->,very thick] (360/9*-\x +90 +10:10mm+\y) -- (360/9*-\x +90-10:10mm+\y)
  \else
     \draw[->,very thick] (360/9*\x +90-10:10mm+\y) -- (360/9*\x +90+10:10mm+\y)
  \fi;
  }
  \end{tikzpicture}
}

\newcommand{\namedconfsix}[3]
{
\begin{tikzpicture}
 \foreach \x in {0,...,5}
  {
  \draw[very thick]  (360/6*\x +90:6mm) circle (2mm) ;
  }

  \foreach \x in #1
    \fill[black]  (360/6*\x +90:6mm) circle (1mm);

 \ifnum #3>-1
   \fill[black] (360/6*#3 +90:6mm) circle (2mm);
 \fi

\foreach \x/\y in #2
    \draw (360/6*\x +90:10mm) node {$\y$};
\end{tikzpicture}
}

\newcommand{\namedconfsixmult}[3]
{
\begin{tikzpicture}
 \foreach \x in {0,...,5}
  {
  \draw[very thick]  (360/6*\x +90:6mm) circle (2mm) ;
  }

  \foreach \x in #1
    \fill[black]  (360/6*\x +90:6mm) circle (1mm);

   \foreach \x in #3
     \fill[black] (360/6*\x +90:6mm) circle (2mm);

\foreach \x/\y in #2
    \draw (360/6*\x +90:10mm) node {$\y$};
\end{tikzpicture}
}

\newcommand\tab[1][.5cm]{\hspace*{#1}}

\newcommand{\gath}{\sc{Gathering}\xspace}
\newcommand{\dgath}{\sc{Distinct Gathering}\xspace}
\newcommand{\gr}{\sc{Gathe}$\mathcal{RR}${\sc{ing}}\xspace}
\newcommand{\fsync}{$\mathcal{FSYNC}$}
\newcommand{\ssy}{$\mathcal{SSYNC}$}
\newcommand{\async}{$\mathcal{ASYNC}$}
\newcommand{\seq}{$\mathcal{SEQ}$}
\newcommand{\rr}{$\mathcal{RR}$}
\newcommand{\gv}{${\check{gv}}$}

\def\BibTeX{{\rm B\kern-.05em{\sc i\kern-.025em b}\kern-.08em
    T\kern-.1667em\lower.7ex\hbox{E}\kern-.125emX}}

\newtheorem{theorem}{Theorem}[section]
\newtheorem{lemma}{Lemma}[section]
\newtheorem{corollary}{Corollary}[section]
\newtheorem{definition}{Definition}[section]

\begin{document}

\title{Oblivious Robots Under Round Robin:\\ Gathering on Rings\\
\thanks{The work has been supported in part by the Italian National Group for Scientific Computation GNCS-INdAM.}
}

\author{\IEEEauthorblockN{
Alfredo Navarra}
\IEEEauthorblockA{\textit{Department of Mathematics and Computer Science} \\
\textit{University of Perugia}\\
Perugia, Italy \\
alfredo.navarra@unipg.it}

\and

\IEEEauthorblockN{
Francesco Piselli}
\IEEEauthorblockA{\textit{Department of Mathematics and Computer Science} \\
\textit{University of Perugia}\\
Perugia, Italy \\
francesco.piselli@unifi.it}
}

\maketitle

\begin{abstract}
Robots with very limited capabilities are placed on the vertices of a graph and are required to move toward a single, common vertex, where they remain stationary once they arrive. This task is referred to as the {\gath} problem. 

Most of the research on this topic has focused on feasibility challenges in the \emph{asynchronous} setting, where robots operate independently of each other. A common assumption in these studies is that robots are equipped with \emph{multiplicity detection}, the ability to recognize whether a vertex is occupied by more than one robot. Additionally, initial configurations are often restricted to ensure that no vertex hosts more than one robot.
A key difficulty arises from the possible symmetries in the robots’ placement relative to the graph's topology. 

This paper investigates the {\gath} problem on Rings under a \emph{sequential} scheduler, where only one robot at a time is active. While this sequential activation helps to break symmetries, we remove two common assumptions: robots do not have multiplicity detection, and in initial configurations, vertices can be occupied by multiplicities.

We prove that such a generalized {\gath} problem cannot be solved under general sequential schedulers. However, we provide a complete characterization of the problem when a sequential \emph{Round Robin} scheduler is used, where robots are activated one at a time in a fixed cyclic order that repeats indefinitely. Furthermore, we fully characterize the {\dgath} problem, the most used variant of {\gath}, in which the initial configurations do not admit multiplicities.

\end{abstract}


\section{Introduction}\label{sec:intro}
In the field of theoretical computer science, swarm robotics is one of the most investigated research areas. 
Robots are usually mobile units with full autonomy that, by operating individually, are able to 
establish some sort of collective behavior in order to solve a common problem. Robots are considered 
in the abstract with their capabilities induced by an underlying model. Those capabilities are usually 
reduced to the minimum, in order to have a more flexible and fault-resistant model. In this context, 
some representative models are, for example, the Amoebot  \cite{DDGRSS14} 
and the Silbot \cite{DDDNP20,NP23,NP23opodis,NP24sss}. One of the most investigated models for a theoretical perspective
in swarm robotics is certainly the $\mathcal{OBLOT}$  \cite{FPS12}. 
In this model, 
robots operate by executing {\tt Look-Compute-Move} cycles. In each cycle, a robot obtains a snapshot 
of the system ({\tt Look}), executes its algorithm to determine the destination of its next movement ({\tt Compute}), and moves toward the computed destination ({\tt Move}).

Within such a context, one of the most popular problems is the so-called {\gath} where robots, 
placed on the vertices of an anonymous graph, are required to reach a common vertex (not known in 
advance) from where they do not move anymore. 

Apart for some impossibility results or basic conditions that guarantee the resolution of the {\gath} problem
provided in \cite{CDN20a, DN17a}, most of the literature usually focuses on specific topologies that are
very symmetric, where the vertices can be partitioned into a few classes of equivalence. Since robots have few topological properties to exploit, the design of a resolution algorithm becomes more challenging.
Those topologies are: 
Trees \cite{DDKN12,DDN13}, Regular Bipartite graphs~\cite{GP13}, Finite Grids \cite{DDKN12}, 
Infinite Grids \cite{DN17}, 
Tori \cite{KLOTW21}, Oriented Hypercubes \cite{BKAS18}, Complete graphs~\cite{CDN20a, CDN19c}, 
Complete Bipartite graphs \cite{CDN20a, CDN19c}, Butterflies \cite{CDDN24}, 
and Rings \cite{KKN10, 
DNN17, DN17a, DDFN18}.	

In most of those studies, the robots operate under an {\emph{asynchronous}} scheduler, where robots are 
activated independently of each other. Other works concern  {\emph{synchronous}} schedulers, where the robots 
share a common notion of time and a subset of activated robots execute their {\tt Look-Compute-Move} cycle at the same time. A very common assumption 
is to have robots 
endowed with the {\emph{multiplicity detection}}. With this property, robots are able to recognize whether 
a vertex contains a {\emph{multiplicity}}, i.e., if two or more robots are located at the same vertex. 

Focusing on rings, these are vertex-transitive graphs where the robots' movements depend entirely
on their relative positioning. So far, on rings, the {\gath} problem has been studied without considering multiplicities in initial configurations. Recently, in~\cite{FW24}, this version of the problem has been referred to as  {\dgath}, and it has been studied for robots moving on the Euclidean plane under a {\emph{Round Robin}} scheduler. This is a specific type of {\emph{sequential}} scheduler, where robots are activated one at a time, in a fixed 
periodic order. The more generic {\emph{sequential}} scheduler, which requires only to activate one robot 
at a time, has been used in \cite{FNPPS2024} to solve the {\emph{Universal Pattern Formation}} ({\tt UPF}) problem. In {\tt UPF}, robots can start from configurations containing multiplicities, and the requirement is to move so as to form a given pattern.

\subsection{Our Results}\label{sec:result}

In this paper, we focus on robots moving on rings and operating under the $\mathcal{OBLOT}$ model with no additional assumptions. We provide an impossibility result for the {\gath} problem under a general \emph{sequential} scheduler, and we present a full characterization for both {\gath} and {\dgath} under the {\emph{round robin}} 
scheduler, proposing an asymptotically time optimal algorithm.

\subsection{Outline}\label{sec:outline}
In the next section, we present and formalize the robot model and the scheduler used. In Section 
\ref{sec:problem}, we formalize the studied problem and present some impossibility results. 
In Section \ref{sec:algo}, 
we present our algorithm to solve {\gath} on rings. In Section \ref{sec:correct}, we present the correctness proof and 
the complexity for the proposed algorithm. 
In Section \ref{sec:runex}, we present an example of execution of the proposed algorithm. 
Finally, in Section \ref{sec:conclusion}, we provide 
concluding remarks and some interesting directions for future works. 

\section{Robot Model}\label{sec:model}

We consider the standard  $\mathcal {OBLOT}$ model of 
distributed systems of  autonomous mobile  robots. In the $\mathcal {OBLOT}$ model, the system is composed of a set  
 $\mathcal{R} = \{r_1, r_2, \dots, r_k\}$ of
  computational {\emph robots} that live and 
  operate on a $n$-vertices anonymous \emph{ring} without orientation.
 We refer to a maximal subset of consecutive empty vertices of the ring as a {\emph{hole}, whereas, we refer to a maximal subset of consecutive occupied vertices as an {\emph{island}.
 Each vertex of the ring is initially empty, occupied by one robot, or occupied by more than one robot (i.e., a \emph{multiplicity}).
 
Robots can be characterized 
  according to many different settings. In particular, they have the following basic properties: 
  
  \begin{itemize}
      \item \textbf{Anonymous:} they have no unique identifiers;
      \item \textbf{Autonomous:} they operate without a centralized control;
      \item \textbf{Dimensionless:} they are viewed as points, i.e., they have no volume nor occupancy restraints;
      \item \textbf{Disoriented:} they have no common sense of orientation;
      \item \textbf{Oblivious:} they have no memory of past events;
      \item \textbf{Homogeneous:} they all execute the same deterministic algorithm with no type of randomization admitted;
      \item \textbf{Silent:} they have no means of direct communication.
  \end{itemize}
   
Each robot in the system has sensory capabilities, allowing it to determine the location of
other robots in the ring, relative to its location. Each robot refers to a {\tt Local Reference System}
({\tt LRS}) that might differ from robot to robot. 
Each robot has a specific behavior described according to the sequence of the following four states: {\tt Wait}, {\tt Look}, {\tt Compute}, and {\tt Move}. Such a sequence defines the computational activation cycle (or simply a cycle) of a robot. More in detail:

\begin{enumerate}
    \item {\tt Wait:} the robot is in an idle state and cannot remain as such indefinitely;
    \item {\tt Look:} the robot obtains a snapshot of the system containing the positions of the other robots with respect to its {\tt LRS}, by activating its sensors. Each robot is seen as a point in the graph occupying a vertex;
    \item {\tt Compute:} the robot executes a local computation according to a deterministic algorithm $\mathcal{A}$ (we also say that the robot executes 
    $\mathcal{A}$). This algorithm is the same for all the robots and its result is the destination of the movement of the robot. Such a destination is either 
    the vertex where the robot is already located, or a neighboring vertex at one hop distance (i.e., only one edge per move can be traversed); 
    \item {\tt Move:} if the computed destination is a neighboring vertex, the robot moves to such a vertex. Otherwise, it executes a {\emph{nil}} 
    movement (i.e., it does not move).
\end{enumerate}

In the literature, the computational cycle is simply referred to as {\tt Look-Compute-Move} ({\tt LCM}) cycle, because when a robot is in the {\tt Wait} state, we say that it is 
\emph{inactive}. Thus, the {\tt LCM} cycle only refers to the \emph{active} states of a 
robot. 
It is also important to notice that since the robots are oblivious, without 
memory of past events, every decision they make during the {\tt Compute} phase is 
based on what they are able to determine during the current {\tt LCM} cycle. In 
particular, during the {\tt Look} phase, the robots take a snapshot of the system and 
they use it to elaborate the information, building what is called the \emph{view} of the 
robot. 
Regarding the {\tt Move} phase of the robots, the movements executed are always considered to be instantaneous. Thus, the robots are only able to perceive the 
other robots positioned on the vertices of the graph, never while moving. Regarding the 
position of a robot on a vertex, it may happen that two or more robots are located on 
the same vertex, i.e., they constitute a multiplicity.

Another important feature that can greatly affect the computational power of the robots is the \emph{time scheduler}. We say that an \emph{epoch}, is the minimum time window within which each robot has been activated at least once.
In general, three main schedulers are used:

\begin{itemize}
    \item \emph{Semi-Synchronous} (\ssy): the activations of the robots are logically divided in global rounds. In each round, one or more robots are activated and obtain the same snapshot. Then, based on the information acquired from the snapshot, they compute and execute their move, completing their cycle by the next round;
    
    \item \emph{Fully-Synchronous} (\fsync): all the robots are activated in every round, executing their {\tt LCM} cycle in a synchronized way;
    
    \item \emph{Asynchronous} (\async): the robots are activated independently of each other and the duration of each phase of the {\tt LCM} cycle is finite but unpredictable. In this scheduler, robots have no common notion of time. Thus, their decisions can be based on obsolete observations of the system.
\end{itemize}

In the {\fsync} case, a round coincides with one epoch.
In the {\ssy} and {\async} cases, it is assumed the 
existence of an \emph{adversary} which determines the computational cycle's timing and which robot(s) 
will be activated. This timing is assumed to be 
\emph{fair}, that is, each robot is able to 
execute its {\tt LCM} cycle within finite time 
and infinitely often. Without this fairness assumption, the adversary could prevent some robot from ever being activated. The duration of an epoch is then finite but unpredictable.

In this work, we consider another type of scheduler: 

\begin{itemize}
    \item \emph{Sequential} (\seq): the robots are activated one at a time, fairness is guaranteed.
\end{itemize}

In particular, we focus on the so-called: 
\begin{itemize}
    \item \emph{Round Robin} (\rr): the robots are activated one at a time in a predetermined order which repeats forever. Each robot is then activated exactly 
once in each epoch. Of course, {\rr} $\subset$ {\seq}, and an epoch equals $k$ rounds, with $k$ being the number of robots in the system.
\end{itemize}

\section{Problem Formulation and Impossibility Results}\label{sec:problem}
The problem we aim to solve is the {\gath} on a $n$-ring, and it is defined as follows: 

\begin{definition}[\gath]\label{def:gathgen}
    Given $k$ robots $r_1, r_2, \dots, r_k$ arbitrarily placed on a $n$-ring, it is required to reach a configuration in a finite number of epochs where 
    exactly one vertex is occupied and from thereon no robot moves.
\end{definition}

We distinguish the general case from the one usually adopted in the literature where initial configurations do not admit multiplicities, that is: 

\begin{definition}[\dgath]\label{def:gathdist}
    Given $k$ robots $r_1, r_2, \dots, r_k$ on a $n$-ring with $k\leq n$, where each vertex is occupied 
    by at most one robot,
    it is required to solve the {\gath}.
\end{definition}

We now define all the cases where {\gath} or {\dgath} are unsolvable. First of all, the next lemma provides a useful property that will be exploited later for both impossibility results and for the designing of the proposed algorithm.

\begin{lemma}\label{lem:movetwo}
    Let $C$ be a configuration on a $n$-ring with exactly $2$ robots placed on different vertices. Under \seq, the only reasonable direction where a robot can move in order to solve the {\gath},
    is toward the other one. 
\end{lemma}

\begin{proof}
    By contradiction, we assume that there exists a gathering algorithm where the movement executed by the robots increases their respective distance, 
    i.e., they move away from each other. To solve the {\gath}, the robots must be located at the same vertex. Thus, since the aim of the executed movement is to increase the distance between the two robots, they will never get close to each other, that is, {\gath} cannot be finalized.
\end{proof}

Note that, as a consequence of this result, even when considering a configuration with more than $2$ robots 
occupying exactly two vertices, since the robots are not able to detect multiplicities, the only 
reasonable movement remains the one toward the other occupied vertex.

We now prove an impossibility result for general schedulers in {\seq}.

\begin{theorem}\label{th:impseq}
    {\gath} on rings using a {\seq} scheduler is impossible for $k$ robots, with $k \geq 3$.
\end{theorem}

\begin{proof}
    Let us consider the stage just before the problem resolution, i.e., when all the 
    robots are placed on two neighboring vertices, say $v_1$ and $v_2$. 
    
    In the case of only $3$ robots $r_1$, $r_2$, and $r_3$, they form a multiplicity with two robots on one vertex, say $r_1$ and $r_2$ are on $v_1$, and the third robot $r_3$ is on $v_2$ (see Fig. \ref{fig:rings1}.a). At this point, 
    the only admissible move to complete the {\gath}, as proven in Lemma \ref{lem:movetwo}, is to move toward the neighboring vertex occupied by robots. 
    Since for a generic scheduler in {\seq} 
    the only condition (apart for fairness) is that only one robot at a time is activated, the 
    adversary can easily activate $r_1$ at time $t$. After $r_1$'s movement, at time $t' > t$, $v_2$ is occupied by a multiplicity. Then the adversary activates robot $r_3$ on $v_2$ and after its movement, at time $t'' > t'$, $v_1$ is again occupied by a multiplicity, and the adversary can activate $r_2$, which moves from $v_1$ to $v_2$. This sequence of activations then starts again, making the resolution of {\gath} impossible. 
    
    In the case of $k > 3$, a similar sequence can be applied by the adversary to activate all the robots fairly.
\end{proof}

Following the above theorem, we choose to work using the {\rr} scheduler, which operates activating all the 
robots sequentially in each epoch, one per round, always maintaining the same sequence. 

We now define the configurations from which solving the {\gath} under {\rr} is impossible for any algorithm.

\begin{definition}[\emph{Unsolvable Configuration}]
    Given a configuration $C$ and a {\rr} scheduler, $C$ is said to be  \emph{Unsolvable} if, for any algorithm $\mathcal{A}$,
    there exists a sequence of activations imposed by the scheduler, that makes the {\gath} impossible to be solved.
\end{definition}

We denote by $\mathcal{UC}$ the set of unsolvable configurations. In the next theorems, we show which configurations belong to $\mathcal{UC}$ with respect to the {\gath} and the {\dgath} problems.

\begin{figure}
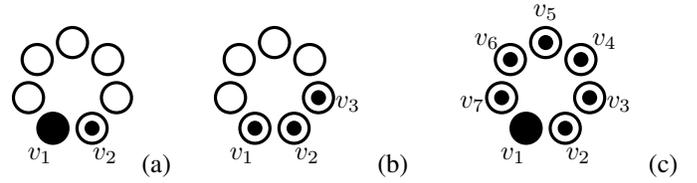

\begin{center}
%
\namedconf{{3,4}}{{3/v_1,4/v_2}}{3}(a)~~~ 
\namedconf{{3,4,5}}{{3/v_1,4/v_2,5/v_3}}{-1}(b)~~~ 
\namedconf{{0,1,2,3,4,5,6}}{{0/v_5,1/v_6,2/v_7,3/v_1,4/v_2,5/v_3,6/v_4}}{3}(c)
\end{center}\caption{Configurations used for the proofs of Theorems \ref{th:impseq} and \ref{th:imprr}. The robots 
are represented by black circles inside vertices. A full black vertex represents a multiplicity. Edges are not drawn for clarity. Labels associated with vertices are used only for analysis purposes: robots are not aware of them as the rings are, in fact, anonymous.}
\label{fig:rings1}
\end{figure}

\begin{theorem}\label{th:imprr}
    Let $C$ be a configuration of $k\ge 3$ robots on a $n$-ring, with: 

    \begin{itemize}
        \item[i)] Only $2$ consecutive vertices occupied;
        \item[ii)] Only $3$ consecutive vertices occupied;
        \item[iii)] All $n$ vertices occupied.
    \end{itemize}

    For each of these types of configurations, there exists a {\rr} scheduler that makes the {\gath} unsolvable.

\end{theorem}

\begin{proof}

    Let us consider a configuration $C$ of type $i)$ with exactly $3$ robots. The impossibility proof is directly inherited from the proof of Theorem 
    \ref{th:impseq}, since the provided scheduler is precisely a {\rr} scheduler.

    Let us consider a configuration $C$ of type $ii)$ with exactly $3$ robots $r_1$, $r_2$, and 
    $r_3$, occupying $3$ consecutive vertices $v_1$, $v_2$, and $v_3$, respectively (see 
    Fig. \ref{fig:rings1}.b).

    By contradiction, let us assume there exists an algorithm $\mathcal{A}$ that solves {\gath} starting from a configuration $C$, composed of $3$ robots occupying $3$ distinct 
    vertices. Whatever $\mathcal{A}$ dictates, 
    the robots must reach a configuration where just two neighboring vertices are occupied before finalizing the {\gath}. 
    
    Let $t$ be the last time in the execution of $\mathcal{A}$ in which the $3$ robots occupy $3$ different vertices. 
    Therefore, since at time $t+1$, a multiplicity composed of $2$ robots is created, necessarily 
    the $2$ robots composing the multiplicity at time $t+1$ must have been neighbors at time $t$. 
    It follows that the possible configurations at time $t$ are:
    $(a)$ $3$ consecutive vertices occupied;
    $(b)$ $2$ neighboring vertices occupied with the third occupied vertex separated by a hole of size greater than $1$;
    $(c)$ $2$ neighboring vertices occupied with the third occupied vertex separated by a hole of size $1$. 

    In a configuration of type $(a)$, let us consider the three robots to be $r_1$ at $v_1$, $r_2$ at 
    $v_2$, and $r_3$ 
    at $v_3$. The possible movements to create a multiplicity consist in: 
    \begin{itemize}
\item[$m'$:] make $r_2$ move toward $v_1$ or $v_3$; 
\item[$m''$:] make $r_1$ or $r_3$ move toward $v_2$. 
     \end{itemize}
   
    If $\mathcal{A}$ applies $m'$, since $r_2$ does not distinguish 
    the two neighbors, the adversary makes $r_2$ move toward the 
    next activated robot at time $t+1$, e.g., $r_1$. At time $t+2$, $r_1$ and $r_2$ are at $v_1$, and 
    $r_3$ is at $v_3$. Now, $r_1$ gets activated, and we know from Lemma \ref{lem:movetwo} that from 
    a configuration with only two occupied vertices, the only available move is to go toward 
    the other occupied vertex. Hence, $r_1$ moves toward $v_3$ creating again a configuration with 
    $3$ consecutive occupied vertices, contradicting the hypothesis that $t$ was the last time during the execution of $\mathcal{A}$ in which $3$ robots occupy $3$ different vertices. Thus, $m'$ is not a feasible move.

    If $\mathcal{A}$ applies $m''$, let us consider the sequence of activations $r_1,r_2,r_3$. After the first 
    epoch, we obtain a configuration with two neighboring vertices, $v_1$ and $v_2$, occupied but with 
    the robots activated alternatively from the two vertices. 
    Again from Lemma \ref{lem:movetwo}, we know that from such a configuration 
    the robots can only move toward the other 
    occupied vertex. From such a configuration and with such a sequence of activations, solving the 
    {\gath} is impossible.\footnote{Note that, with the sequence of activations $r_1,r_3,r_2$, in the same instance, {\gath} would be solved.}

    Let us consider now a configuration of type $(b)$ with $r_1$ at $v_1$, $r_2$ at $v_2$, and 
    $r_3$ at distance greater than $2$ from $v_2$. 
    Recalling that, by hypothesis, $t$ was the last time of the execution of 
    $\mathcal{A}$ in which $3$ different vertices were occupied, the only feasible movement at $t+1$
    consists in making $r_1$ move toward $v_2$, or making $r_2$ move toward $v_1$. 
    After one of such moves, only $r_3$ should move at any time 
    to avoid creating a configuration with $3$ different vertices occupied, but this  cannot be forced since any activated robot would 
    see the same configuration composed of only two occupied vertices. Therefore, a configuration of 
    type $(b)$ cannot be the last configuration with $3$ different vertices occupied by $3$ robots.

    Finally, let us consider a configuration of type $(c)$, with $r_1$ at $v_1$, $r_2$ at $v_2$, and 
    $r_3$ at $v_4$ at distance $2$ from $v_2$.
    In order to not increase the size of the hole, falling into the previously described case $(b)$, the only feasible movement consists in making 
    $r_1$ move toward $v_2$. Subsequently, if $r_2$ is activated, then by Lemma~\ref{lem:movetwo}, again $3$ different vertices become occupied, contradicting the hypothesis about time $t$.\footnote{Note that if $r_3$, instead of $r_2$, is the next robot activated by the scheduler, then {\gath} would be solved.}

    Summarizing, we have proven so far that in order to solve the {\gath} problem, algorithm $\mathcal{A}$ should somehow force a specific \rr\ scheduler. We now show that starting from the configuration $C$ with $r_1$ at $v_1$, $r_2$ at $v_2$, and 
    $r_3$ at $v_3$, it is not possible for $\mathcal{A}$ to force such a scheduler, contradicting the hypothesis that $\mathcal{A}$ solves the {\gath}.
    
    Let us consider again the initial configuration $C$, and all the possible movements that make possible 
    reaching a 
    configuration from which solving the {\gath} is possible. 
    
   Configuration $C$ is the same as the one considered in case $(a)$. On the one hand, 
    we know that executing move $m'$ surely does not lead to the resolution of the problem. 
    On the other hand, executing $m''$ could lead to the resolution of the problem, but only with an 
    appropriate scheduler. Since $C$ is our initial configuration, we can consider to have the 
    scheduler $r_1,r_2,r_3$, which does not lead to the resolution of the problem because the 
    robots will be activated alternatively from two consecutive vertices.

    We do not need to consider reaching the configuration of case $(b)$ starting from $C$, 
    because we have already seen that from such a configuration {\gath} is unsolvable.

    Finally, to reach a configuration of type $(c)$ from $C$, one robot between $r_1$ and $r_3$ must 
    have moved to separate itself from the other two. Let us consider the scheduler $r_3,r_1,r_2$. 
    To reach such a configuration, $r_3$ moves toward $v_4$. Then, in order to maintain a 
    configuration with a hole of size $1$ separating the islands of robots (as configurations with larger holes have been proven to be 
    unsolvable), the next robot to move is $r_1$, which moves toward $v_2$. At this point, $r_3$ is at $v_4$, and $r_1$ and $r_2$ are at 
    $v_2$. The last robot to move in the first epoch is $r_2$, moving from $v_2$ to $v_3$ 
    according to Lemma \ref{lem:movetwo}. Therefore, 
    after one epoch, the configuration is again composed of $3$ robots occupying $3$ consecutive vertices, i.e., those movements only lead to a loop.

    From all of the obtained results, we can state that starting from a configuration $C$ with $3$ consecutive vertices occupied, it is not 
    possible to force a specific \rr\ scheduler under which the hypothetical algorithm $\mathcal{A}$ would solve the {\gath}.
    
    Finally, we consider a configuration $C$ of type $iii)$, i.e., when the ring is completely occupied by robots. Let us consider the case with $k=n+1$ robots, i.e., 
    there is exactly one multiplicity. Let us call the vertices in the ring $v_1,v_2,\dots,v_n$, let the multiplicity be $v_1$, occupied by $r_1$ 
    and $r_2$, and the vertices $v_2,\dots,v_n$ occupied by robots $r_3,\dots,r_{n+1}$, respectively (see Fig. \ref{fig:rings1}.c). 
    Let us consider the sequence of activations $r_2, r_3, \dots, r_{n+1}, r_1$. 
    With such a configuration the only possible move is to go toward one of the two occupied neighbors. Since the robots are not able to distinguish 
    their neighbors, the adversary makes them move all in counterclockwise order, i.e., from $v_1$ to $v_2$, from $v_2$ to $v_3,\dots$, from $v_n$ to $v_1$. 
    With this configuration, scheduling and movements, the activated robot $r_i$ is always on a multiplicity.
    Therefore, the movement executed by each robot, just transfers the multiplicity from one vertex to 
    another, thus the configuration always remains 
    with $n$ occupied vertices and {\gath} is unsolvable.

    It is worth noting that the provided proofs of cases $i)$, $ii)$, and $iii)$, can be easily extended to any $k>3$, $k>3$, and $k>n+1$, respectively. 
\end{proof}

Other configurations belonging to $\mathcal{UC}$ are presented in the following results.

\begin{theorem}\label{th:3/5}
 Let $C$ be a configuration on  a $5$-ring
 where $k\geq5$ robots occupy $3$ vertices.
 If a vertex occupied by a multiplicity is neighboring to another occupied vertex, then the {\gath} problem is unsolvable from $C$ under a {\rr} scheduler.
\end{theorem}

\begin{proof}
    The proof proceeds by providing a sampling configuration $C$ composed of $k=5$ robots occupying $3$ distinct vertices, and exhaustively considering all the possible movements. 
    
   About different positioning of the robots or about the case $k>5$, similar arguments can be deduced.
    
    From the proof of Theorem \ref{th:imprr}, we already know that if the $3$ occupied vertices are consecutive, then the {\gath} is unsolvable. 
    Therefore, we consider the only other possible positioning of robots on $3$ vertices on a $5$-ring, i.e., an occupied vertex $v_1$ with two empty 
    neighbors $v_2$ and $v_5$, and the other two vertices, $v_3$ and $v_4$, occupied (see, for instance, Fig. \ref{fig:rings2}.a). 
    
    Let us consider a configuration with exactly $k=5$ robots, with one robot $r_1$ 
    located at $v_1$, robots $r_2$ and $r_3$ located at $v_3$, and robots $r_4$ and $r_5$ located at $v_4$. Now, we have to consider 
    all the possible movements executed by the robots: 
    $a)$ the robot at $v_1$ moves toward one of its two empty neighbors; 
    $b)$ the robots positioned 
    on $v_3$ and $v_4$ move toward their occupied neighbor; 
    $c)$ a robot among those positioned on $v_3$ or $v_4$ moves toward its empty neighbor. 

    \begin{itemize}
        \item Considering case $a)$, 
        if $r_1$ is the first robot to move toward one of its neighbors, say $v_5$, 
        after such a movement the configuration is composed of $3$ consecutive vertices occupied. Let the sequence of 
        activations be $r_1,r_2,r_3,r_4,r_5$.
        By Theorem \ref{th:imprr} and the movements described therein, with these assumptions, {\gath} is unsolvable;
    
        \item Considering case $b)$, let the sequence of activations be $r_1,r_2,r_4,r_3,r_5$. Since the robots at $v_3$ and $v_4$ are activated 
        alternatively, the occupied vertices never change, hence {\gath} is unsolvable;

        \item Considering case $c)$, let 
        $r_2$ be the first robot to move from $v_3$ to $v_2$. After such a movement,
        a configuration with $4$ consecutive vertices occupied is generated. 
        
        Now, if the next movement dictates to move the external robots at $v_1$ or $v_4$ toward their empty neighbor, then the configuration does not change, 
        maintaining $4$ consecutive vertices occupied or, reaching a configuration with all $5$ vertices occupied. In the latter case, a robot can only 
        move toward one of its occupied neighbors, reaching again a configuration with $4$ consecutive vertices occupied. Since no other movements are
        allowed, this loop makes the {\gath} unsolvable.

        If, instead, from the configuration with $4$ consecutive vertices occupied, the external robots $r_1$ at $v_1$, and $r_4$ and $r_5$ at $v_4$, have to move 
        toward their occupied neighbor, if $r_1$ is the first robot to move, such a movement creates 
        a configuration with only $3$ consecutive vertices occupied. According to the movements
        described in the proof of Theorem \ref{th:imprr}, we know that from such a configuration 
        {\gath} is unsolvable. If, instead, a different sequence of activations dictates 
        that $r_4$ and $r_5$ are the first robots to move, after their movement again a configuration 
        with only $3$ consecutive vertices occupied is generated, i.e., {\gath} is unsolvable.

        The last possible movement with $4$ consecutive vertices, consists in making 
        the robots at $v_2$ and $v_3$, move toward each other or toward 
        $v_1$ and $v_4$, respectively. In both cases, the generated configuration has only $3$ 
        vertices occupied divided in two islands, one of size $1$ and the other of size $2$. Now, 
        since we are in case $c)$, the robots from the island of size $2$ move toward their 
        empty neighbor, thus creating again the same configuration with $3$ vertices occupied. 
        The described movements are only able to generate a configuration with two islands of size $1$ and $2$, respectively, or a configuration with $4$ vertices occupied, hence {\gath} is unsolvable.

    \end{itemize}  
    \end{proof}

    Details on different positioning of the robots or for $k>5$ are omitted but can be easily obtained by the provided arguments. Note
    that, according to the scenario provided in the above proof, in a configuration where 
    there is only one multiplicity located at $v_1$ and two single robots at $v_3$ and $v_4$, respectively, it is possible to find a solution for the problem.

\begin{figure}[t]
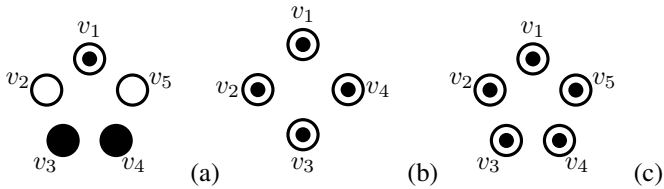

\begin{center}
\namedconffivemult{{0,2,3}}{{0/v_1,1/v_2,2/v_3,3/v_4,4/v_5}}{{2,3}}(a) 
\hspace{-5mm}
\namedconffour{{0,1,2,3}}{{0/v_1,1/v_2,2/v_3,3/v_4}}{-1}(b) 
\hspace{-3mm}
\namedconffive{{0,1,2,3,4}}{{0/v_1,1/v_2,2/v_3,3/v_4,4/v_5}}{-1}(c) 
\end{center}\caption{Configurations used for the proofs of Theorem \ref{th:3/5}, Lemma \ref{lem:dist4/4} and Lemma \ref{lem:dist5/5}, respectively.}\label{fig:rings2}
\end{figure}

\begin{corollary}\label{coro:4/5}
 Let $C$ be a configuration of $k\geq5$ robots occupying $4$ vertices of a $5$-ring. The {\gath} problem is
    unsolvable from $C$ under a {\rr} scheduler.
\end{corollary}

\begin{proof}
    The proof simply follows by observing that in the proof of Theorem \ref{th:3/5}, {\gath} has been proved to be unsolvable also for the case of $k=5$ robots on a $5$-ring occupying 
    $4$ distinct vertices, given a specific {\rr} scheduler.    \end{proof}

All the configurations addressed by Theorem \ref{th:imprr}, Theorem \ref{th:3/5}, and Corollary 
\ref{coro:4/5}, compose 
the set $\mathcal{UC}$ of unsolvable configurations for the {\gath} problem.

Concerning the {\dgath} problem, the proof of case $ii)$ of Theorem \ref{th:imprr} still 
holds, and the only other unsolvable configurations we detected are given by $4$ robots on a $4$-ring and by $5$ robots on a $5$-ring. 

\begin{lemma}\label{lem:dist4/4}
    Let $C$ be a configuration of $4$ robots occupying all vertices of a $4$-ring. The {\dgath} problem is
    unsolvable from $C$ under a {\rr} scheduler.
\end{lemma}

\begin{proof}
Let $r_1$, $r_2$, $r_3$, and $r_4$, be the $4$ robots in $C$ occupying vertices $v_1$, $v_2$, $v_3$, $v_4$, respectively, on a $4$-ring, (see Fig. \ref{fig:rings2}.b).
Starting from $C$, the only movement that any robot can execute is to go toward one of its two neighbors, say $r_1$ moves toward $v_4$. After this movement, the configuration has $3$ consecutive vertices occupied. By Theorem 
\ref{th:imprr}, we have that the obtained configuration is unsolvable, hence $C$ is as such. In fact, it is sufficient to choose the sequence $r_1$, $r_4$, $r_2$, $r_3$ as activation scheduler. The movements dictated by the proof of Theorem~\ref{th:imprr} would prevent the robots to ever accomplish the {\gath}. 
\end{proof}

\begin{lemma}\label{lem:dist5/5}
    Let $C$ be a configuration of $5$ robots occupying all the nodes of a $5$-ring. The {\dgath} problem is
    unsolvable from $C$ under a {\rr} scheduler.
\end{lemma}

\begin{proof}
    When the ring is completely occupied (see Fig. \ref{fig:rings2}.c), the only movement that any robot can execute once activated is to go toward one of its occupied neighbors. After such 
    a movement, the ring is composed of $1$ empty vertex and $4$ occupied vertices, with one of those being a multiplicity. 
    In the proof of Theorem \ref{th:3/5}, we already proven that there exists a {\rr} scheduler that makes such a configuration unsolvable with respect to the {\gath} problem. Thus, the statement holds.
\end{proof}

The set $\mathcal{UC}$ 
for the {\dgath} problem is then
composed of any $n$-ring with exactly $3$ consecutive vertices occupied by $3$ robots
and by the configurations composed of a $4$-ring or a $5$-ring fully occupied.

According to the obtained impossibility results, in the next section we are going to fully characterize both the {\gath} and the  {\dgath} problems. 
We provide a unique resolution algorithm 
that works for any configuration $C \notin \mathcal{UC}$ for both problems.

\begin{table}[t]
\bgroup
\def\arraystretch{1.4}
\setlength{\tabcolsep}{6pt}
\begin{center}
  \begin{tabular}{ | c | p{0.85 \columnwidth}| }
    \hline
    \textit{Var}  &  \textit{Definition}  
                  \\ \hline \hline
               
    \texttt{$\mathtt{b4}$}  & 
        There is exactly one hole constituted of at most $4$ vertices
               \\ \hline

    \texttt{$\mathtt{b5}$}  &
        There is exactly one hole constituted of at least $5$ vertices
               \\ \hline

    \texttt{$\mathtt{f}$}  & 
        There are no holes
               \\ \hline
               
    \texttt{$\mathtt{h}$}  & 
        There are exactly two holes, one constituted of just $1$ vertex, the other of more than $1$ vertex
               \\ \hline
               
    \texttt{$\mathtt{o1}$}   & 
        All robots occupy exactly $1$ vertex ({\gath} accomplished)
               \\ \hline
               
    \texttt{$\mathtt{o2}$} & 
        All robots occupy exactly $2$ neighboring vertices 
               \\ \hline
    \texttt{$\mathtt{o3}$}  & 
        All robots occupy exactly $3$ consecutive vertices
               \\ \hline
    \texttt{$\mathtt{p}$} & 
        All robots occupy exactly $2$ vertices separated by $1$ empty vertex 
               \\ \hline

  \end{tabular}
\end{center}
\egroup
\caption{ The basic boolean variables used to define all the tasks' preconditions. }
\label{tab:basic-variables2}
\end{table}
%

\begin{table}[t]
\bgroup
\def\arraystretch{1.4}
\setlength{\tabcolsep}{6pt}
\begin{center}
  \begin{tabular}{ | l | l | l | r | l | l |}
    \hline
\multicolumn{2}{|l|}{\textit{Sub-Problem}} & \textit{Task} &  
       \textit{Precondition} & \textit{Move} & \textit{Transitions}\\ \hline \hline


\multicolumn{2}{|l|}{\multirow{3}{*}{\raggedleft $I_{\check{gv}}$ }} & \textit{ $T_1$} & $true$  &  $m_1$  &  $T_1, T_2, T_4, T_5$ \\ \cline{3-6}

\multicolumn{2}{|l|}{}  &  \textit{ $T_2$}  &  $\mathtt{b4} \vee \mathtt{f}$  &  $m_2$ &$T_1, T_2, T_4$ \\ \cline{3-6}

\multicolumn{2}{|l|}{}  &  \textit{ $T_3$}  &  $\mathtt{b5}$  &  $m_3$ &$T_2,T_3, T_4$ \\ \cline{1-6}

\multicolumn{2}{|l|}{ $N_{\check{gv}}$ }  &  \textit{ $T_4$}  &  $\mathtt{h}$  &  $m_4$  &  $T_4, T_5$ \\ \cline{1-6}

\multicolumn{2}{|l|}{\multirow{2}{*}{\raggedleft $O_{\check{gv}}$ }}  &  \textit{ $T_5$}  &  $\mathtt{p}$  &  $m_5$ & $T_6, T_7$ \\ \cline{3-6}

\multicolumn{2}{|l|}{}  &  \textit{ $T_6$}  &  $\mathtt{o3}$  &  $m_6$  &  $T_6, T_7$ \\ \cline{1-6}

\multicolumn{2}{|l|}{$Finalize$}  &  \textit{ $T_7$}  &   $\mathtt{o2}$  &  $m_7$  &  $T_7, T_8$ \\ \cline{1-6}

\multicolumn{2}{|l|}{ $Term$ }  &  \textit{ $T_8$}  &  $\mathtt{o1}$  &  $nil$  &  $T_8$ \\

\hline
  \end{tabular}
\end{center}
\egroup
\caption{Schematization of Algorithm {\gr} for solving both the {\gath} and the {\dgath} problems.}
\label{tab:tasks2}
\end{table}

\begin{table}[t]
\bgroup
\def\arraystretch{1.4}
\setlength{\tabcolsep}{6pt}
\begin{center}
  \begin{tabular}{|c| p{0.41 \textwidth}|}
    \hline
    \textit{Move}  &  \textit{Description}  
                  \\ \hline \hline

    $m_1$  & 
    $if$ $r$ is neighboring one of the biggest holes, $then$ \newline
        \tab $if$ all the islands are of size $2$, $then$ \newline
            \tab \tab $if$ there is no unique biggest hole, $then$ \newline
                \tab \tab \tab $r$ moves toward its closest empty vertex \newline
            \tab \tab $else$ $r$ moves toward its occupied neighbor \newline
        \tab $else$ $if$ $n-2$ vertices are occupied but not consecutive, $then$ \newline
            \tab \tab \tab $if$ $r$ is neighboring only one empty vertex, $then$ \newline
                   \tab \tab \tab \tab $r$ moves toward its occupied neighbor \newline
            \tab \tab $else$ $r$ moves away from the biggest hole
               \\ \hline
    $m_2$  & $if$ both neighbors of $r$ are occupied, $then$ \newline
                \tab  $if$ $n=6$ and the unique hole has size $1$, $then$ \newline
                    \tab \tab $if$ $r$ is not on the farthest vertex from the  hole, $then$ \newline
                        \tab \tab \tab $r$ moves away from the hole \newline
                \tab $else$ $r$ moves toward the hole if any, or toward any direction
               \\ \hline
    $m_3$  & $if$ $r$ admits an empty neighbor $x$, $then$\newline 
             \tab $r$ moves toward $x$
               \\ \hline
    $m_4$  & $if$ $r$ is neighboring the biggest hole and one robot, $then$\newline
             \tab $r$ moves toward its occupied neighbor
               \\ \hline    
    $m_5$  & $r$ moves toward the other occupied vertex
               \\ \hline
    $m_6$  & $if$ $r$ is neighboring an empty vertex, $then$\newline 
             \tab $r$ moves toward its occupied neighbor
               \\ \hline
    $m_7$  & $r$ moves toward its occupied neighbor
               \\ \hline               
  \end{tabular}
\end{center}
\egroup
\caption{ Description of the moves from the point of view of a robot $r$. }
\label{tab:moves}
\end{table}
%

\section{\gath on Rings}\label{sec:algo}

The algorithm presented in this paper is designed according to the methodology proposed in \cite{CDN21a}. Let us now briefly summarize how an algorithm $\mathcal{A}$, conceived to solve a generic problem $\mathcal{P}$, can be designed using that methodology, 

Recall that we are considering a model where the robots have very weak capabilities: they can only wake up, take a 
snapshot of the graph, and based on that observation they can take a deterministic decision. For those reasons, it is 
better to consider the problem $\mathcal{P}$ as composed of a series of sub-problems such that, each sub-problem, is 
simple enough to be solved by a ``task'' executed by one or more robots. Therefore, let us assume that the problem 
$\mathcal{P}$ is decomposed into simple tasks $T_1, T_2, \dots, T_q$, where one of them is the \emph{terminal} one, 
i.e., the one where the robots recognize that the problem $\mathcal{P}$ is solved and they do not execute any other move.

As we previously described, the robots operate following the {\tt LCM} cycle, hence, they must be able to recognize which 
task they have to execute according to the configuration that they sense during the {\tt Look} phase. This recognition 
can be executed by providing $\mathcal{A}$ with a predicate $P_i$ for each task $T_i$. Once a robot wakes up and 
perceives that a certain predicate $P_i$ is true, following $\mathcal{A}$ it knows that the task $T_i$ must be 
executed in order to solve a sub-problem. Going into more detail, with predicates well-formed, algorithm 
$\mathcal{A}$ can be used in the {\tt Compute} phase as follows: if once awakened, a robot $r$ executing algorithm $\mathcal{A}$, detects that a certain predicate $P_i$ is true, then $r$ executes a move $m_i$ associated with the task $T_i$. 
For this approach to be valid, each well-formed predicate must guarantee the following properties:

\begin{itemize}
    
    \item $Prop_1$: each predicate $P_i$ must be computable on the configuration $C$ sensed by the robot in the {\tt Look} 
    phase; 
    
    \item$Prop_2$: $P_i$ $\wedge$ $P_j =$ {\tt false}, for each $i \neq j$; thanks to this property the robots are able 
    to precisely recognize which task to execute, without ambiguity;
    
    \item$Prop_3$: for each possible configuration $C$ sensed, there must exist a predicate $P_i$ evaluated true by an 
    activated robot.
    
\end{itemize}

To be recognized by the robots, each task $T_i$ requires some \emph{precondition} to be verified. Hence, for 
the definition of the predicates $P_i$, we need to define some \emph{basic variables} that capture metric/numerical/ordinal/topological aspects of the configuration $C$     
        sensed by the robots, that can be evaluated by each activated robot, based solely on the observations made during
        the {\tt Look} phase.

Let us assume that $\tt pre_i$ is the composition of the variables characterizing the preconditions 
of $T_i$, for each $1\leq i \leq q$. The predicate $P_i$ can then be defined as follows: 
\begin{equation}\label{eq:predicate}
    P_i = {\tt pre_i} \wedge \lnot({\tt pre_{i+1}} \vee {\tt pre_{i+2}} \vee \dots \vee {\tt pre_q})
\end{equation}
With this definition, we are sure that any predicate satisfies property $Prop_2$.

Let us now consider an execution of algorithm $\mathcal{A}$, where a task $T_i$ is executed with respect to the current
configuration $C$. The configuration $C'$, generated by $\mathcal{A}$ after the execution of task $T_i$, must be 
assigned to a task $T_j$. Then, we can say that algorithm $\mathcal{A}$ can generate a transition from $T_i$ to $T_j$. 
The set of all possible transitions of $\mathcal{A}$ determines a directed graph called \emph{transition graph}. Note 
that the terminal task among $T_1, T_2, \dots, T_q$, where the problem $\mathcal{P}$ is solved, must be a sink vertex in the transition graph.

In \cite{CDN21a}, it is shown that the correctness of an algorithm $\mathcal{A}$ designed with the proposed methodology, can be 
obtained by proving that all the following properties hold:

\begin{itemize}
    \item[$H_1$:] The transition graph is correct, i.e., for each task $T_i$, the tasks reachable from $T_i$ by means of 
        transitions are exactly those represented in the transition graph;
    \item[$H_2$:] Apart for the self-loop induced by a terminal task, all the other loops in the transition graph, 
        including self-loops, must be executed a finite number of times;
    \item[$H_3$:] With respect to the studied problem $\mathcal{P}$, no unsolvable configuration is generated by 
        $\mathcal{A}$.
\end{itemize}


\subsection{High Level Description of Algorithm {\gr}}\label{sec:description}

The proposed Algorithm {\gr}, is designed according to the methodology recalled previously. Following such an approach, the main problem {\gath} is subdivided into a set of 
sub-problems. 
Solving all the sub-problems, leads the 
algorithm to solve the {\gath}. 

The first sub-problem that needs to be solved by the robots consists in having a configuration where 
the robots occupy exactly two distinct islands with only one empty vertex separating 
them on one side, and more than one empty vertex separating them on the other side. 
Creating this configuration allows the robots to uniquely identify the same vertex where to solve the {\gath}. This vertex is the empty (single) one 
separating the two islands. 
In the following, such a vertex will be called \emph{gathering vertex} and denoted by
{\gv}. 

A different sub-problem consists in moving all the robots toward the two vertices 
neighboring {\gv}, until only those two vertices are occupied. From that point, the next sub-problem requires to have only two neighboring 
vertices occupied by robots. To reach such a configuration, 
all the activated robots move toward 
{\gv} until the only two vertices occupied by robots are {\gv} 
itself and one of its neighbors. Finally, the last sub-problem makes the robots move 
toward {\gv} until that is the only vertex occupied by robots, i.e., 
{\gath} is solved.

\subsection{Description of Sub-problems and Tasks}\label{sec:sptasks}

In this section, we describe all the details of the five sub-problems we defined and the designed tasks to solve each of them. The basic variables we used to define the corresponding preconditions are shown in Table~\ref{tab:basic-variables2}, whereas Table~\ref{tab:tasks2} shows the corresponding preconditions and transitions.
For each task, 
the movements described are the ones presented in Table~\ref{tab:moves}.

The first sub-problem $I_{\check{gv}}$ consists in creating a specific configuration where the robots are divided in two 
islands separated on one side by a hole of size $1$, and on the other side by a hole of size at least 
$2$. This configuration allows the robots to uniquely identify the gathering vertex {\gv}, hence the name 
$I_{\check{gv}}$. To solve this sub-problem we 
need the robots to operate following the designed tasks. For $I_{\check{gv}}$, three tasks, $T_1$, $T_2$ and $T_3$, are possibly needed. 

\medskip

\noindent {\tt Task $T_1$}. Task $T_1$ activates when the following predicate holds:

\begin{adjustwidth}{15pt}{}
{$P_1 =$} $({\tt pre_{1}} \equiv  true) \wedge \lnot({\tt pre_{2}} \wedge {\tt pre_{3}} \wedge \dots \wedge {\tt pre_8})$ \\
\end{adjustwidth}
According to the defined preconditions, see Table~\ref{tab:tasks2}, it follows that the configurations addressed by this task are those admitting at least two holes, and 
if they have exactly two holes, those holes are both of unitary size or both of size greater than one.
In this case, the scheduled robot $r$ 
wakes up and executes move $m_1$.

Only a robot neighboring to one of the biggest holes can move. There are various cases that can occur. In particular, if the islands are all of size $2$ and there is no unique biggest hole, then $r$ moves toward its closest empty vertex.  

In a different scenario managed by task $T_1$, if $n-2$ vertices are occupied and not consecutive,
 $r$ can move if it is neighboring to only one empty vertex;
the executed movement is directed toward the occupied neighbor. 

In any other case, any robot neighboring the biggest 
hole will move to the opposite direction with respect to such a hole with the goal of increasing its size.

Note that, if there are multiple holes with the same
biggest size, they are all considered as the ``biggest hole''.

\medskip

\noindent {\tt Task $T_2$}. Task $T_2$ is activated when there is exactly one hole in the graph of size at most $4$, or when there is no hole
at all. In the latter case, according to Theorem~\ref{th:imprr}, of course only {\dgath} can be solved. The following predicate holds: \\
\begin{adjustwidth}{15pt}{}
{$P_2 =$} $({\tt pre_{2}} \equiv  \mathtt{b4} \vee \mathtt{f}) \wedge \lnot({\tt pre_{3}} \wedge {\tt pre_{4}} \wedge \dots \wedge {\tt pre_8})$ \\
\end{adjustwidth}
If the activated robot $r$ is not neighboring to an empty vertex, 
then it is able to move executing $m_2$. In this case, $r$ moves toward the closest empty hole or,
in the case of no holes, it moves toward any direction. 
The goal of this movement is to 
create a new hole of size $1$.

\medskip

\noindent {\tt Task $T_3$}. In task $T_3$, the goal of the algorithm is to create a configuration where, 
starting with all the robots positioned consecutively, and a unique hole of size at least $5$, 
there are two different islands  
of robots, where the smallest hole separating them is of size $1$. In particular, task $T_3$ 
activates when the following predicate holds: \\
\begin{adjustwidth}{15pt}{}
{$P_3 =$} $({\tt pre_{3}} \equiv  \mathtt{b5}) \wedge \lnot({\tt pre_{4}} \wedge {\tt pre_{5}} \wedge \dots \wedge {\tt pre_8})$ \\
\end{adjustwidth}
An activated robot $r$ moves executing move $m_3$ if it is neighboring to an empty vertex $x$. The movement is directed toward $x$, attempting to create 
a new hole of size $1$.

The subsequent sub-problem requires to create a configuration where there are exactly two vertices occupied by the robots, separated by a hole 
of size $1$ on one side, and a hole of size at least $2$ on the other side. Focusing on the two occupied vertices neighboring {\gv}, we call this 
sub-problem $N_{\check{gv}}$.
This sub-problem requires the execution of task $T_4$.

\medskip

\noindent {\tt Task $T_4$}. When the robots form exactly two islands with the smallest hole 
separating them of size $1$ and the other one of size at least $2$, task $T_4$ activates. In particular, the predicate that holds is the 
following: \\
\begin{adjustwidth}{15pt}{}
{$P_4 =$} $({\tt pre_{4}} \equiv  \mathtt{h}) \wedge \lnot({\tt pre_{5}} \wedge {\tt pre_{6}} \wedge \dots \wedge {\tt pre_8})$ \\
\end{adjustwidth}
The goal of this task is to position all the robots on exactly two vertices, with one empty vertex in-between them. 
To achieve such a configuration, an activated robot $r$ executes move $m_4$: if $r$ is neighboring a robot and 
an empty vertex 
which is part of the biggest hole in the graph, then $r$ moves toward its occupied neighbor. 
With such 
a movement, the islands gradually reduce their size until only two vertices are occupied.

Once $N_{\check{gv}}$ has been solved, the subsequent sub-problem consists in starting to occupy the gathering vertex {\gv}, until 
only that vertex and one of its neighbors are occupied, hence the name $O_{\check{gv}}$. 
This sub-problem possibly 
requires the execution of tasks $T_5$ and $T_6$. Note that, until the 
start of those two tasks, vertex {\gv} is empty which is fundamental for the correct resolution of the {\gath}.

\medskip

\noindent {\tt Task $T_5$}. Task $T_5$ is activated when all the robots are neighboring {\gv}, with {\gv} still empty. Hence, the following 
predicate holds: \\
\begin{adjustwidth}{15pt}{}
{$P_5 =$} $({\tt pre_{5}} \equiv  \mathtt{p}) \wedge \lnot({\tt pre_{6}} \wedge {\tt pre_{7}} \wedge \dots \wedge {\tt pre_8})$ \\
\end{adjustwidth}
In this case, the activated robot $r$ moves in the direction of the other occupied vertex executing move $m_5$. In doing so, $r$ becomes the 
first robot to occupy {\gv}.

\medskip

\noindent {\tt Task $T_6$}. After a robot moved according to move $m_5$ and {\gv} becomes occupied, if $3$ consecutive vertices are occupied with 
{\gv} being the middle one, then task $T_6$ is executed and the following predicate holds: 
\\
\begin{adjustwidth}{15pt}{}
{$P_6 =$} $({\tt pre_{6}} \equiv  \mathtt{o3}) \wedge \lnot({\tt pre_{7}} \wedge {\tt pre_{8}})$ \\
\end{adjustwidth}
During this task, only the external robots neighboring {\gv} are able to move executing move $m_6$: once
 activated, those robots move toward 
{\gv}, moving forward with the {\gath} onto that vertex.

At the completion of $O_{\check{gv}}$, only two neighboring vertices are occupied, with one being what was uniquely identified by the 
robots as the gathering vertex {\gv}. At this point, the two occupied vertices are indistinguishable but, since all the previous movements 
executed during tasks $T_5$ and $T_6$ moved the robots onto {\gv}, only the robots positioned on {\gv}'s neighbor still have to be activated 
by the adversary within the same epoch. 
Therefore, this sub-problem is the \emph{finalization} one, denoted by \emph{Finalize}, which requires the execution of task $T_7$.

\medskip

\noindent {\tt Task $T_7$}. Once only two neighboring vertices are occupied, task $T_7$ is activated and the following predicate holds:
\\
\begin{adjustwidth}{15pt}{}
{$P_7 =$} $({\tt pre_{7}} \equiv  \mathtt{o2}) \wedge \lnot({\tt pre_{8}})$ \\
\end{adjustwidth}
The goal of this task is to move all the robots on the same vertex to solve the {\gath}. Hence, the only move that can be 
executed by the robots is to move toward their occupied neighbor, i.e., once activated they 
execute move $m_7$.

Once the \emph{Finalize} sub-problem has been solved, there is nothing else to do, hence we called the last sub-problem \emph{Term}, since 
it is the termination stage of the algorithm. The task associated with this sub-problem is task $T_8$.

\medskip

\noindent {\tt Task $T_8$}. When the {\gath} has been solved, all the robots occupy the same vertex and task $T_8$ activates, with the following predicate holding:
\\
\begin{adjustwidth}{15pt}{}
{$P_8 =$} $({\tt pre_{8}} \equiv  \mathtt{o1})$ \\
\end{adjustwidth}
During this task, the activated robots recognize that the problem has been solved. Thus, they only execute the \emph{nil} movement that does not change the configuration.

\begin{figure}[ht]
    \centering
    \includegraphics[width=.9\linewidth]{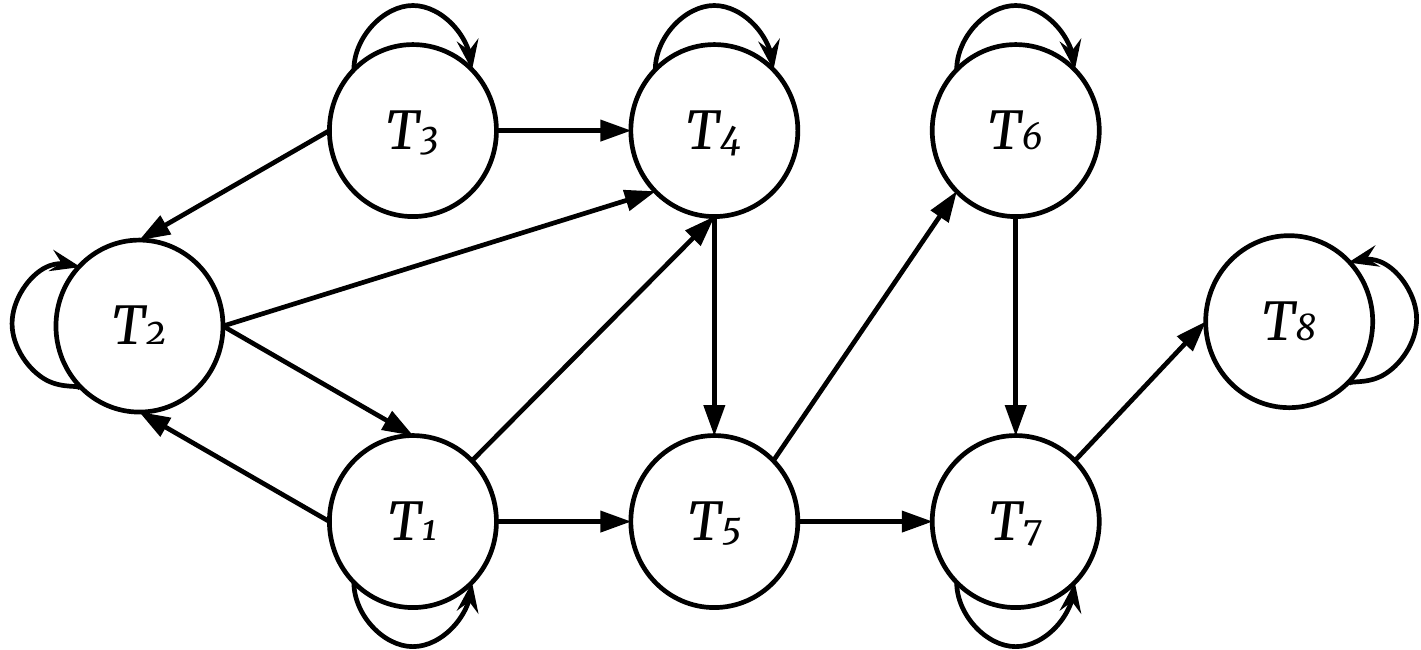}
    \caption{Transition graph derived from Table \ref{tab:tasks2}.}
    \label{fig:transitions}
\end{figure}

\section{Correctness}\label{sec:correct}
The predicates used in the algorithm are well-formed since they guarantee that the properties $Prop_1$, $Prop_2$ and $Prop_3$ introduced in Section 
\ref{sec:algo}, are all valid. In particular, $Prop_1$ follows from the preconditions presented in Table \ref{tab:tasks2}; $Prop_2$ is valid because 
each predicate $P_i$ has been defined according to Eq. \ref{eq:predicate}; $Prop_3$ follows from the definitions of the
predicates $P_i$, e.g., if $P_8$, $P_7, \dots,P_2$ are all false, then $P_1$ holds. 

We first provide
a specific lemma for each task where we show that the 
properties $H_1$, $H_2$, and $H_3$, introduced in Section \ref{sec:algo}, hold. 
Then, a final theorem, will combine all the lemmata to prove the correctness of {\gr}.

\begin{lemma}\label{lem:t1}
    Let $C$ be a solvable configuration in $T_1$. From $C$, in less than $n$ epochs, {\gr} leads to a configuration belonging to $T_2$, $T_4$, or $T_5$.
\end{lemma}

\begin{proof} 

    In task $T_1$, Algorithm {\gr} selects a robot to move with the general goal of increasing the 
    size of the biggest hole in the system.  \smallskip
    
    \noindent $H_1$: 
        If there are exactly two holes of size $1$ in the ring, 
        once a 
        robot $r$ neighboring a hole is activated at time $t$, it moves toward its occupied neighbor at time $t'>t$. If at 
        time $t$ robot $r$ was on a multiplicity, then at $t'$ the configuration looks unchanged to the robots, and hence 
        it is still in $T_1$; otherwise, if $r$ was not on a multiplicity, one hole of size $1$ at 
        time $t$ becomes a hole of size $2$ at $t'$. 
        Thus, predicate $P_4$ becomes true and the configuration is in $T_4$ after at most one epoch. 
        
        A special case of two holes of size $1$ can happen in a $6$-ring where there are two islands, both of size $2$. When this happens, the activated robot $r$, neighboring a hole, moves toward its closest hole. If $r$ 
        was not on a multiplicity before moving, then after the move it creates a new configuration still in $T_1$ but with an island of size $3$ and one of size $1$. If, instead, $r$ was on a multiplicity, by $r$'s movement, a configuration 
        with one single hole of size $1$ is created, which is in $T_2$. 

        In the case of rings with more than 6 nodes but 
        admitting only islands of size 2, similarly to the special case of the 6-ring specified before, robots move toward empty vertices, filling the holes and enlarging some island (with the configuration either in $T_4$ or still in $T_1$).
     
        In the case of no holes of size $1$ or when the islands are not all of size $2$, 
        only the robots neighboring any biggest hole can move toward the opposite direction with respect to such a hole. Once one of those robots moves, if it was on a multiplicity, 
        then the new configuration looks unchanged to the robots, and hence it is still in $T_1$. 
        If, instead, the moving robot 
        was not on a multiplicity, the biggest hole is enlarged. Hence, in subsequent epochs, either the configuration is still in $T_1$ but with one biggest hole, or it may fall in $T_4$ or $T_5$. 

        The case of $n-2$ vertices not occupied with two islands, one of which of size $1$, constitutes a special case where the robot(s) composing the island of size $1$ are prevented to move in order to avoid to reach a configuration in $T_2$.
        \smallskip

        \noindent $H_2$: 
        A robot $r$ moves only if it is neighboring one of the biggest holes in the system. Robot $r$ moves toward the hole either if it is 
        neighboring two different biggest holes, or if all the islands are of size $2$.

        In the latter case, after one movement, at least one island has changed its size.
        
        In the former case, after one movement, $r$ will ``leave behind'' a hole of size greater than before, moving 
        toward a hole of smaller size. Thus, this movement has a specific direction. Since the scheduler only activates one robot at a time, each 
        change in the configuration will be detected by the next activated robots. Hence, once a unique biggest hole is created, its neighboring robots will move to gradually 
        increase its size until there are only two holes in the system and the smallest has size $1$, i.e., the configuration is in $T_4$ or $T_5$. 
        
        The maximum number of epochs of this self-loop, happens when initially there are only islands of size $2$ and all the 
        holes have the same size. 
        In such a case, one move is used to change the size of one island, creating a new one of size $3$, or splitting one island into two of size 
        $1$ both. 
        After the first movement, the islands are not all of size $2$ anymore, hence the robots neighboring a biggest 
        hole can move in the opposite direction in order to increase its size. Moreover, since a hole has now 
        its size decreased, at least one island is neighboring one of the biggest holes on one side and 
        a smaller hole on the other side. Thus, by moving away from the biggest hole, at 
        the end of the first epoch, for sure one unique biggest hole is created.
        
        In each subsequent epoch, the unique biggest hole will increase its size of $2$ until one robot's movement 
        creates a configuration with exactly two holes, with one of them of size $1$, and the other of 
        greater size.
        Therefore, in less than $n$ epochs, a configuration in $T_4$ or $T_5$ is reached.
        \smallskip

        \noindent $H_3$: 
        As the intent of move $m_1$ is in general to enlarge the biggest hole and leave at least another hole, 
        no configuration with a unique island of size $2$, $3$, or $n$, can be generated from $T_1$ (cf. Theorem~\ref{th:imprr}). Consequently, also a configuration with $3$ or $4$ vertices occupied of a $5$-ring (cf. Theorem~\ref{th:3/5} and Corollary~\ref{coro:4/5}, respectively) is not reachable.
\end{proof}

\begin{lemma}\label{lem:t2}
    Let $C$ be a solvable configuration in $T_2$. From $C$, in at most $1$ epoch, {\gr} leads to a configuration belonging to $T_1$ or $T_4$.
\end{lemma}

\begin{proof}
    In task $T_2$, Algorithm {\gr} selects a robot $r$ to move, only if it is neighboring two occupied vertices. The goal of this 
    task is to create one or two holes. \smallskip
        
        \noindent $H_1$: 
        In $T_2$ the configuration has no holes or only one hole of size at most $4$. In the former case, multiplicities are not considered since 
        that would be a configuration in $\mathcal{UC}$. Therefore, once a robot moves, it creates a hole of size $1$, and the configuration is still 
        in $T_2$. 
        
        Consider now a configuration already admitting a hole. Apart for the special case of a $6$-ring, once a robot $r$ with 
        two occupied neighbors decides to move, it moves toward the hole. If $r$ was not on a multiplicity, it creates a new hole of size $1$, hence in one epoch the configuration is 
        in $T_1$. Instead, if $r$ was on a multiplicity, the configuration does look the same, remaining in $T_2$. However, since the movements are directed toward the hole, all the robots occupying the farthest vertex from the hole are ensured to leave such a vertex within one epoch. Once a new hole of size $1$ is created, the configuration admits two holes. If both holes have size $1$, the obtained configuration belongs to $T_1$, otherwise it is in $T_4$. 
        
        When $C$ concerns a $6$-ring with exactly one hole of size $1$, move $m_2$ is slightly different as it concerns robots occupying two specific vertices of the ring. Anyway, in one epoch, it ensures to create a new hole of size $1$ and the obtained configuration is in $T_1$.
        \smallskip

        \noindent $H_2$: 
        If there is a unique hole in the configuration, the movement of robot $r$ is always 
        directed toward a specific direction. 
        Hence, the number of movements while the configuration remains in $T_2$ is limited by the number of robots involved in the movements. In any case, in one epoch if a robot moves, all the robots occupying the same vertex move, thus creating a new hole of size $1$. 
        In case of no holes, as already mentioned, there are no multiplicities, hence there is no self-loop and after one move (hence one epoch) 
        the configuration concerns the previous case. 
        \smallskip

        \noindent $H_3$: 
        The goal of task $T_2$ is to create a configuration with two holes. In doing so,
        no configuration with a unique island of size $2$, $3$, or $n$ is generated from $T_2$ (cf. Theorem~\ref{th:imprr}). 
        Therefore, also a configuration with $3$ or $4$ vertices occupied on a $5$-ring (cf. Theorem~\ref{th:3/5} and Corollary~\ref{coro:4/5}, respectively) is not reachable.
\end{proof}

\begin{lemma}\label{lem:t3}
    Let $C$ be a configuration in $T_3$. From $C$, in less than $2$ epochs, {\gr} leads to a configuration belonging to $T_2$ or $T_4$.
\end{lemma}

\begin{proof}
    In task $T_3$, Algorithm {\gr} selects to move one of the most external robots of the unique island (i.e., the robots with one 
    empty neighbor), with the goal of creating a new hole in the configuration. \smallskip

        \noindent $H_1$: 
        In $T_3$, the configuration has a unique hole of size $j \geq 5$. Let us consider a hole of size exactly $5$. Once a robot $r$ with one 
        empty neighbor is activated, such a robot moves toward the empty neighbor. If $r$ was on a multiplicity, then the configuration now has only one hole 
        of size $4$, hence in one epoch predicate $P_2$ holds and the configuration is in $T_2$. 
        Instead, in the case of $r$ not being part of a multiplicity, once it moves, 
        it creates a new hole of size $1$, while reducing the other hole. Hence in one epoch the configuration has two holes, one of size $1$ and the other of 
        size $4$. Thus, predicate $P_4$ holds and the configuration is in $T_4$.

        In the case of the unique hole of size $j \geq 6$, if before moving $r$ was on a multiplicity, after its movement the configuration continues to 
        have a unique hole but of size $j-1$. Hence, the configuration is still in $T_3$. Instead, if $r$ was not on a multiplicity, in 
        one epoch a new hole of size $1$ is created and the configuration is in $T_4$.
        \smallskip

        \noindent $H_2$: 
        The self-loops only happen when the unique hole has size $j \geq 6$. Let us consider a hole of size $j=7$. Once a robot $r$ part of a 
        multiplicity completes its movement, the unique hole has its size reduced to $6$. In the same epoch, another robot $r'$ is selected to move, 
        from ``the other side'' of the island. Let us say that $r'$ was also part of a multiplicity. Again, 
        after $r'$ completes its movement, 
        the configuration maintains a unique hole 
        and its size is $5$. Now, the next robot able to move once activated is $r$, which is for sure not part of a multiplicity. Hence, after it 
        moves, it creates a new hole and the configuration is not in $T_3$ anymore. Therefore, in at 
        most $2$ epochs, the obtained configuration is either in $T_2$ or $T_4$. Note that the same 
        arguments also hold for any hole of size $j>7$.
        \smallskip

        \noindent $H_3$: 
        In task $T_3$, there is a hole of size at least $5$. Therefore, the configurations of Theorem~\ref{th:3/5} and Corollary~\ref{coro:4/5} are not considered, since they refer to a $5$-ring. Moreover, the goal of task $T_3$ 
        is to create a hole of size $1$ inside the unique island, thus obtaining two different islands. Therefore, with such a movement, 
        no configuration with a unique island of size $2$, $3$, or $n$, can be reached from $T_3$ (cf. Theorem~\ref{th:imprr}).        
\end{proof}

\begin{lemma}\label{lem:t4}
    Let $C$ be a configuration in $T_4$. From $C$, in at most $n-5$ epochs {\gr} leads to a configuration belonging to $T_5$.
\end{lemma}

\begin{proof}
    In task $T_4$, Algorithm {\gr} selects to move a robot $r$ that, on one side, is neighboring to the biggest hole of $C$, and on the other 
    side, to an occupied vertex. The movement executed by $r$ is directed toward its occupied neighbor. The 
    goal of 
    such a movement is to reduce the size of the island which 
    $r$ is part of, until there are 
    exactly two occupied vertices, separated by a hole of size $1$. \smallskip

        \noindent $H_1$: 
        In task $T_4$, there are exactly two islands, separated by one hole of size $1$ and one hole of size greater than $1$. The robot $r$ that can move occupies one of the vertices of the borders of an island of size at least $2$. Once $r$ moves, if it was part of a multiplicity, then the configuration appears unchanged. Otherwise, 
        the size of the island of which $r$ is part of is reduced of $1$. Once the size of both islands becomes $1$, the configuration 
        is composed of exactly two vertices occupied by the robots, with a hole of size $1$ separating them on one side. 
        Thus, predicate $P_5$ holds and the 
        configuration is in $T_5$. 
        \smallskip

        \noindent $H_2$: 
        The movement of a robot $r$ always has a specific direction, i.e., toward the hole of size $1$. Each movement reduces the size of an island, or 
        reduces the number of robots on the multiplicity occupying one vertex at the border of an island. To exit the self-loop, both islands must have size $1$. 
        Let us consider the maximum size of an island.
        Such size is given by $n$ minus the sum of: the size of the other 
        hole in the system which is at least $2$; the size of the other island which is at least $1$; the hole of size $1$ separating the two islands. Overall, we obtain $n-4$.
        Since we want this biggest island to become of size $1$, at each epoch the island reduces its size of $1$. Thus, in at most 
        $n-5$ epochs, such a result is accomplished and the configuration is in $T_5$.
        \smallskip

        \noindent $H_3$: 
        The goal of task $T_4$ is to increase the size of the biggest hole until the robots are positioned on exactly two vertices (i.e., 
        two islands of size $1$) separated by a hole of size $1$. Therefore,
        no configuration with a unique island of size $2$, $3$, or $n$, is generated from $T_4$ (cf. Theorem~\ref{th:imprr}). Moreover, 
        during such movements, also a configuration with $3$ or $4$ vertices occupied of a $5$-ring (cf. Theorem~\ref{th:3/5} and Corollary~\ref{coro:4/5}, respectively) cannot be reached.
\end{proof}

\begin{lemma}\label{lem:t5}
    Let $C$ be a configuration in $T_5$. From $C$, in one movement, {\gr} leads to a configuration belonging to $T_6$ or $T_7$.
\end{lemma}

\begin{proof}
    In task $T_5$, algorithm {\gr} selects to move a robot $r$  with the goal of 
    creating a configuration with only $2$ or $3$ consecutive vertices occupied.
    \smallskip

        \noindent $H_1$: 
        Let us call the two occupied vertices $v_1$ and $v_3$, separated by an empty vertex $v_2$. Once a robot $r$ moves, say from $v_1$ to $v_2$, only two 
        different situations can happen depending on whether $v_1$ contains a multiplicity or not.
        If $v_1$ does not contain a multiplicity, after $r$'s movement,  
        only the vertices $v_2$ and $v_3$ are occupied. Therefore, 
        predicate $P_7$ holds and the configuration is in $T_7$. 
        Otherwise, if $r$ was on a multiplicity, again after $r$'s movement, all $3$ vertices $v_1$, $v_2$, and $v_3$, become 
        occupied. Thus, predicate $P_6$ holds and the configuration is in $T_6$. \smallskip

        \noindent $H_2$: 
        There are no self-loops from $T_5$ since after one move, the configuration has a unique island 
        of size $2$ or $3$. \smallskip

        \noindent $H_3$: 
        The goal of task $T_5$ is to create a configuration where there is a unique island with size $2$ or $3$. The move executed 
        during this task, defines the first movement with which a robot occupies the gathering vertex {\gv}.
        Note that, for a configuration to be in $\mathcal{UC}$, it is not enough to have a unique island of size $2$ or $3$. 
        Together 
        with that, a specific {\rr} scheduler is needed to make the problem unsolvable. From $T_5$, the configuration 
        with only $2$ or $3$ consecutive vertices occupied can be obtained, as described before, by making one robot $r$ move from $v_1$ (or $v_3$). After $r$'s 
        movement, before such a robot is activated again in a new epoch, all the other robots in the configuration will be activated. Hence, as it will be described later in the proof of task $T_7$, 
        they will all move toward $v_2$, that is, {\gv}.
\end{proof}

\begin{lemma}\label{lem:t6}
    Let $C$ be a configuration in $T_6$ obtained from $T_5$. From $C$, in at most $1$ epoch, {\gr} leads to a configuration belonging to $T_7$.
\end{lemma}

\begin{proof}
     In task $T_6$, only $3$ consecutive vertices $v_1$, $v_2$, and $v_3$, are occupied. The robots selected by Algorithm {\gr} to move are the ones 
     located at $v_1$ or $v_3$. Their movement is directed toward $v_2$. Furthermore, since $C$ has been obtained from $T_5$, the last robot that moved is the one at $v_2$, i.e., $v_2$ does not contain a multiplicity.\smallskip

        \noindent $H_1$: 
        Let us say that the robot $r$ selected to move is located at $v_1$. If $v_1$ contains a multiplicity, then after $r$'s movement the configuration 
        looks the same. Otherwise, only the vertices $v_2$ and $v_3$ will be occupied, hence predicate $P_7$ holds and the configuration is in $T_7$. In any 
        case, a configuration in $T_7$ is reached in at most one epoch since all the robots at $v_1$ and $v_3$ get activated by the scheduler and move 
        toward $v_2$. Hence the first vertex among $v_1$ and $v_3$ that gets empty leads the configuration to $T_7$.
        \smallskip

        \noindent $H_2$: 
        The number of movements that give the self-loop in $T_6$ are limited by the number of robots located at $v_1$ and $v_3$. 
        The activated robots at $v_1$ and $v_3$ will continue to move toward $v_2$ until one of such two vertices becomes empty.
        Hence, in at most one epoch, 
        the obtained configuration is in $T_7$. \smallskip

        \noindent $H_3$: 
        The goal of task $T_6$ is to reach a configuration with a unique island of size $2$.
        Before reaching $C$, by Lemma~\ref{lem:t5}, 
        a robot $r'$ moved from $v_1$ or $v_3$ toward $v_2$. All the robots that are activated after that 
        movement and before activating $r'$ again, are located at $v_1$ and $v_3$ and in $T_6$ they all would move toward $v_2$. With such movements, 
        once the configuration is composed of only two consecutive vertices occupied, 
        all the robots at $v_2$ have already been activated and the ones 
        located at the other occupied vertex still need to be activated. 
        Therefore, this configuration with such a scheduling is not unsolvable.
\end{proof}

\begin{lemma}\label{lem:t7}
    Let $C$ be a configuration in $T_7$  obtained from $T_5$ or $T_6$. From $C$, in at most $1$ epoch, {\gr} leads to a configuration belonging to $T_8$.
\end{lemma}

\begin{proof}
    In task $T_7$, only two consecutive vertices are occupied. Algorithm {\gr} will make the robots move toward their occupied neighbor. \smallskip

        \noindent $H_1$: 
        By assumption, $C$ has been reached from $T_5$ or from $T_6$. 
        
        At $T_5$, the vertices occupied were $v_1$ and $v_3$, with $v_2$ 
        empty in-between them. Reaching $T_7$ from $T_5$ means that a robot $r$ moved from $v_1$ or $v_3$, toward $v_2$. Let us say that $r$ moved 
        from $v_1$ and now the only two occupied vertices are $v_2$ and $v_3$. Since $r$ at $v_2$ has just moved, all the robots at $v_3$ still need to be 
        activated before activating $r$ again. Hence, they will move toward $v_2$ one by one. The movement executed by a robot $r'$ at $v_3$ does not change 
        the configuration if $v_3$ is a multiplicity, otherwise, $v_2$ becomes the only occupied vertex. Therefore, in at most one 
        epoch, predicate $P_8$ holds and the 
        configuration is in $T_8$, i.e., {\gath} has been solved.

        If, instead, we consider that $C$ has been reached from a configuration $C'$ in $T_6$, we must recall that, according to Lemma~\ref{lem:t6}, $C'$ can be reached only from $T_5$. This means that if the 
        two occupied vertices are now $v_2$ and $v_3$, all the robots occupying $v_2$ are those moved as last from $v_1$ and possibly from $v_3$. 
        Therefore, 
        the robot(s) at $v_3$ still have to be activated before activating again those at $v_2$. Once all the robots from $v_3$  have moved to $v_2$, the configuration is in $T_8$ and the problem is solved. \smallskip

        \noindent $H_2$: 
        As previously described, the robots that move in this task are all located at the same vertex. Hence, the number of self-loops is limited 
        by the number of robots located at such a vertex. In at most one epoch, the reached configuration is in task $T_8$ and   the problem is solved. \smallskip

        \noindent $H_3$: 
        With all the previously described movements, the configuration from which the predicate of $T_7$ is activated for the first time,
        is reached with a scheduling that activates all the robots from one vertex, before activating 
        those located at the other vertex. Thus, this configuration leads directly to the resolution of the problem since all the 
        activated robots will move toward their neighboring occupied vertex.
\end{proof}

\begin{theorem}
    {\gr} solves the {\gath} and the {\dgath} problems for each configuration $C$ not belonging to the corresponding set $\mathcal{UC}$ in at most $n-3$ epochs.
\end{theorem}

\begin{proof}
    Lemmata \ref{lem:t1}-\ref{lem:t7} ensure that properties $H_1$, $H_2$, and $H_3$ hold for each task $T_1, T_2, \dots, T_7$. All the transitions are 
    those reported in Table \ref{tab:tasks2} and represented in Fig. \ref{fig:transitions}. It follows that the only possible loop that must be considered is the one involving $T_1$ and $T_2$. On that matter, by Lemma~\ref{lem:t1} we have that $T_2$ is reachable from $T_1$ only in the specific setting of a $6$-ring, 
    where the occupied vertices are divided in two islands of size $2$. From the obtained configuration in $T_2$ after $m_1$, the next movement brings the configuration back to $T_1$ but with two islands of different size. Hence the edge from $T_2$ to $T_1$ can be traversed at most once. Therefore, the system goes out of the loop between $T_1$ and $T_2$ in finite time.

    The correctness of {\gr} revolves around the fact that once a gathering vertex {\gv} has been identified by the robots (i.e., a configuration in $T_5$ is reached), in exactly one epoch the {\gath} problem is solved.
    In fact, passing through $T_6$ and $T_7$, all the robots move one by one toward {\gv}. Once a configuration in $T_8$ is reached, no robot will move anymore, recognizing that the problem has been solved.

    Regarding the time complexity of the algorithm, it is sufficient to put together the results obtained in Lemmata 
    \ref{lem:t1}-\ref{lem:t7}. First of all, it is worth noting that from $T_5$, in at most $2$ epochs,  
    {\gath} is solved. This is given by the fact that {\gv} has been identified and, at the next activation, 
     each robot will move toward the same vertex. Those activations can be divided in $2$ different 
    epochs depending on when $T_5$ started.
    
    The task requiring the largest amount of epochs is $T_4$. 
    In fact, 
    in $T_1$ the biggest hole increases its size of $2$ in each epoch except of the last one, before 
    obtaining a configuration in $T_4$ or $T_5$. Instead, considering the worst case of $T_4$, the biggest hole increases its 
    size of only $1$, while the biggest island decreases its size of the same amount. Therefore, starting from any configuration in $T_1$, $T_2$, $T_3$, or $T_4$, in at most $n-5$ epochs 
    the configuration is for sure in $T_5$. From there, as stated before, in at most $2$ more epochs the problem is solved. By Theorem~\ref{th:imprr}, initial configurations cannot belong to $T_6$ nor to $T_7$. Thus, 
    Algorithm {\gr} solves the {\gath} problem in at most $n-3$ epochs.
\end{proof}

    It is worth noting that an ideal optimal algorithm in terms of time, brings all the robots toward the same vertex in at least $\left \lfloor{\frac{n}{2}}\right \rfloor$ 
    epochs.
    To see this, by considering a ring fully occupied, regardless of the position of the gathering vertex, 
    there exists a robot which has to travel  a distance of $\left \lfloor{\frac{n}{2}}\right \rfloor$ hops. This can be done in at least $\left \lfloor{\frac{n}{2}}\right \rfloor$ epochs. Therefore, our algorithm is asymptotically optimal.
%

\section{Running Example}\label{sec:runex}

In Fig. \ref{fig:runex}, it is represented a running example of a $6$-ring with $5$ robots and one multiplicity. 
In Fig. \ref{fig:runex}.a the 
configuration starts in $T_1$ with two holes, both of size $1$, and two islands, both of size $2$. 
In Fig. \ref{fig:runex}.b, a robot has moved according to $m_1$ from $v_6$ to $v_1$, bringing the configuration into $T_2$ 
with a unique hole of size $1$. 
In Fig. \ref{fig:runex}.c, after another movement from a robot at $v_6$ to $v_1$, according to $m_2$, the configuration is again in $T_1$ but now the islands are not only of size $2$. 
In Fig. \ref{fig:runex}.d, the robot at $v_3$ has moved according to $m_1$ away from its empty 
neighbor to increase the size of that hole, bringing the configuration into $T_4$ with one hole of size $1$ 
and the other hole of size $2$. Note that from this configuration, the gathering vertex {\gv} has been 
recognized by all the robots as $v_6$ and they will all gather on such a vertex.
In Fig. \ref{fig:runex}.e, the robots at $v_2$ moved toward $v_1$, according to $m_4$, to increase the 
size of the biggest hole, bringing the configuration in $T_5$, with all the robots occupying 
exactly two vertices separated by a hole of size $1$. 
In Fig. \ref{fig:runex}.f, the robots started moving from $v_1$ to 
$v_6$, according to $m_5$, occupying $3$ consecutive vertices and bringing the configuration in $T_6$. 
In Fig. \ref{fig:runex}.g, the robot at  
$v_5$ moved toward $v_6$, according to $m_6$, bringing the configuration in $T_7$, with only two consecutive 
occupied vertices. 
Finally, in Fig. \ref{fig:runex}.h, the robots moved from $v_1$ toward $v_6$, according to $m_7$, since such robots are those that remain to be activated within the epoch that started by the time the configuration reached $T_5$.
By exploiting this peculiarity based on the {\rr} scheduler,  a configuration 
in $T_8$ is reached,  and the {\gath} is solved.

\begin{figure}[t]
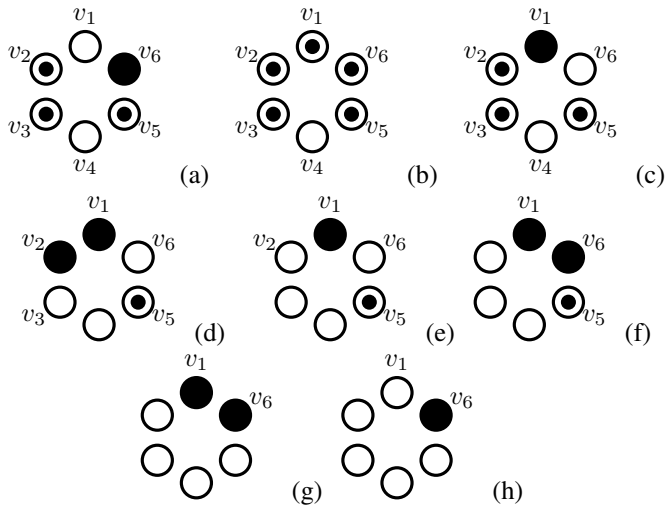

\begin{center}
\namedconfsix{{1,2,4,5}}{{0/v_1,1/v_2,2/v_3,3/v_4,4/v_5,5/v_6}}{5}(a)
\namedconfsix{{0,1,2,4,5}}{{0/v_1,1/v_2,2/v_3,3/v_4,4/v_5,5/v_6}}{-1}(b)
\namedconfsix{{0,1,2,4}}{{0/v_1,1/v_2,2/v_3,3/v_4,4/v_5,5/v_6}}{0}(c)

\namedconfsixmult{{4}}{{0/v_1,1/v_2,2/v_3,4/v_5,5/v_6}}{{0,1}}(d)
\namedconfsixmult{{4}}{{0/v_1,1/v_2,4/v_5,5/v_6}}{{0}}(e)
\namedconfsixmult{{4}}{{0/v_1,4/v_5,5/v_6}}{{0,5}}(f)

\namedconfsixmult{{}}{{0/v_1,5/v_6}}{{0,5}}(g)
\namedconfsixmult{{}}{{0/v_1,5/v_6}}{{5}}(h)

\end{center}
\caption{Running example on a $6$-ring with $5$ robots and one multiplicity at $v_6$.}
\label{fig:runex}
\end{figure}

\section{Conclusion}\label{sec:conclusion}
We have studied the {\gath} problem within rings. In particular, first we have shown that under a 
generic {\emph{sequential}} (\seq) scheduler, {\gath} is unsolvable. 
Then, we focused on solving the problem 
under the {\emph{Round Robin}} (\rr) sequential scheduler, offering a complete characterization and proposing 
a resolution algorithm. The same algorithm also solves the {\dgath} problem when starting from any solvable configuration. 

As a future work, it would be interesting to approach the {\gath} problem, under \rr\ or any other scheduler within {\seq},
also on different graph topologies.

\bibliography{references}





\end{document}